# Lessons Learned Applying Deep Learning Approaches to Forecasting Complex Seasonal Behavior

Andrew T. Karl[1], James Wisnowski[1], Lambros Petropoulos[2]

[1]Adsurgo LLC, Pensacola, FL
[2]USAA, San Antonio, TX

**Abstract**
Deep learning methods have gained popularity in recent years through the media and the relative ease of implementation through open source packages such as Keras. We investigate the applicability of popular recurrent neural networks in forecasting call center volumes at a large financial services company. These series are highly complex with seasonal patterns - between hours of the day, day of the week, and time of the year - in addition to autocorrelation between individual observations. Though we investigate the financial services industry, the recommendations for modeling cyclical nonlinear behavior generalize across all sectors. We explore the optimization of parameter settings and convergence criteria for Elman (simple), Long Short-Term Memory (LTSM), and Gated Recurrent Unit (GRU) RNNs from a practical point of view. A designed experiment using actual call center data across many different "skills" (income call streams) compares performance measured by validation error rates of the best observed RNN configurations against other modern and classical forecasting techniques. We summarize the utility of and considerations required for using deep learning methods in forecasting.
**Key Words:** ARIMA, Time Series

## 1. Introduction

Member contact call centers receive fluctuating call volumes depending on the day of the week, the time of day, holidays, business conditions, and other factors. It is important for call center managers to have accurate predictions of future call volumes in order to manage staffing levels efficiently. The call center arrival process has been well documented and explored in the literature (Gans, Koole, & Mandelbaum, 2003). In the application presented here, there are several different "skills" (or "splits") to which an incoming call may be routed – depending on the capabilities of the call center agents – and an arrival volume forecast is required for each skill in the short term for day-ahead or week-ahead predictions.

The weekly seasonality found in call arrivals can be modeled effectively through a variety of methods to include Winter's Seasonal Smoothing (Winters, 1960) or Autoregressive Integrated Moving Average (Box & Jenkins, 1970). Some accessible references for many of these concepts aimed at the practitioner are well documented in the literature (e.g. Bisgaard & Kulachi, (2007, 2008)) while recommended texts are Bisgaard & Kulachi (2011) and Montgomery, Jennings & Kulachi (2015).

Aiming to improve on these classic methods, "doubly stochastic" linear mixed models (Aldor-Noiman, Feigin, & Mandelbaum, 2009) have effectively modeled additional complexities as outlined in a recent review paper from Ibrahim, Ye, L'Ecuyer, & Shen (2016). Similarly, Recurrent Neural Networks (RNNs) have been recommended as deep learning approaches to forecast call volume for a wireless network (Bianchi et al., 2017) in addition to numerous other applications including ride volumes with Uber (Zhu & Laptev, 2017). While the doubly stochastic and RNN approaches to predicting call volumes offer greater flexibility in modeling complex arrival behavior by incorporating exogenous variables, this flexibility comes at the cost of greater computational and programming complexity (as well as greater prediction variance). This paper explores practical aspects of managing that complexity for these models, applies the models to actual call volumes recorded by a large financial services company, and compares the prediction capability to that of the more traditional Winters smoothing and ARIMA models.

First, we modify computational aspects of the doubly stochastic approach proposed by Aldor-Noiman, Feigin, & Mandelbaum (2009) to improve call center forecasting performance. Doubly stochastic implies a two-level randomization where not only are call arrivals random variables, but also the call arrival mean parameter. Forecasts are produced by taking advantage of the unique correlation structure for each split while accounting for trend, seasonality, cyclical behavior, and serial dependence. The doubly stochastic model is more complex than ordinary regression as it accounts for both inter- and intra-day correlation. We suggest modifications to the originally proposed approach that lead to more stable convergence and more flexible behavior when many splits need to be fit.

Secondly, we consider how RNNs may be used to model incoming call volume. Whereas "traditional" densely connected, feedforward neural networks process each data point independently, RNNs process sequences according to temporal ordering and retain information from previous points in the sequence. As it processes points within sequences, the RNN maintains states that contain information about what it has seen previously in the sequence (Chollet & Allaire, 2018). This intra-sequence memory is useful in time series applications to autocorrelated data. In the context of call center volumes, these sequences could be constructed to correspond to individual days of observations over a fixed number of (e.g. 30 minute) periods. Bianchi et al. (2017) consider three different RNN architectures to model incoming call volume over a mobile phone network: Elman Recurrent Neural Networks (ERNN) (Elman, 1990), Gated Recurrent Units (GRU) (Cho, et al., 2014), and Long Short Term Memory (LSTM) (Hochrieter & Schmidhuber, 1997), listed in order of increasing complexity.

These three RNNs, along with the dense neural network, are now available via the R Keras package (Allaire & Chollet, 2018). Once code has been written for one of the RNNs, the user can switch between the other two by toggling a single option (and, after data reformatting, switch to a dense network). This offers the potential – via a designed experiment – to produce a pragmatic answer to the question of which type of (R)NN provides the best fit to the process at hand. Whereas Bianchi, Maiorino, Kampffmeyer, Rizzi, & Jenssen (2017) created their experimental design by randomly generating points within the design space and then selected the design that lead to the minimum error rate, we create a full factorial design (treating all factors as categorical to allow arbitrary shape in the otherise (discrete) continuous factor of number of nodes) and then explore the behavior of the error rates across the design space with a profiler for the resulitng linear model for the error rate as a function of the NN settings. Unlike ARIMA or regression (including doubly stochastic) modeling approaches for time series, there is a stochastic behavior in the predictions made by neural networks due to the use of randomly initialized weights. Unless the seed for the software's random number generator is fixed, repeated fitting of the same neural network will lead to different predictions. The amount of variation in the resulting predictions depends on the complexity of the network and on the steps that have been taken to avoid overfitting, including early stopping of the optimizer. When selecting a model configuration, we will not only want to minimize the expected error rate, but also minimize the variability in the error rates. To this end, we seek to minimize the upper 95% prediction interval on the testing error rate. The NN study proceeds in two phases where a screening experiment first identifies the most useful (R)NN, followed by a more comprehensive performance study against common forecasting approaches across many more skills.

Section 2 describes the doubly stochastic model for call volumes and how modifications to the originally proposed computational approach can lead to improved convergence. Section 3 details how a full factorial design is used to characterize the performance of RNN options as a function of five factors (and their interactions) on the resulting short-term forecast error rate. Additionally, Section 3 describes the selection of the model configuration that leads to the minimum upper bound on the 95% prediction interval for the testing error rate. Due to the number of different model configurations that must be run along with the computational complexity of RNNs, the first phase discussed in Section 3 considers only a limited number of skills and validation days. In Section 4, the best performing RNNs are run over a larger validation set and over all call center skills to compare the performance to the doubly stochastic mixed model approach, and to ARIMA as well as Winters smoothing.

## 2. Stable Settings for Fitting the Mixed Model

There are two distinct influences on call volumes that induce a correlation between the observed call counts, violating the independence assumption made by ordinary least squares regression models that might be used to model the volumes (Ibrahim & L'Ecuyer, 2013). Within a given day, some event may lead to more/fewer calls than expected. For example, unexpected behavior in the stock market in the morning may lead to an increased number of calls for the rest of the day at a financial services contact center. This is intra-day correlation. Likewise, there are systemic processes responsible for inter-day correlation. Heuristically, if we noticed that the residuals are very large and positive throughout the day today caused by a weather event for example, we might also expect a larger-than-average call load tomorrow. Ignoring correlation between subsequent observations leads to inaccurate standard errors and prediction intervals. In addition, although the estimates from a linear regression may be unbiased in the presence of correlated residuals, they will not be efficient (Demidenko, 2013).

It is typical for call center regression models to include a day-of-week by period-of-day interaction (Ibrahim, Ye, L'Ecuyer, & Shen, 2016). In a call center open five days per week with 32 half-hour periods per day, this interaction involves 160 fixed-effect parameters. In addition, a call center may require forecasting for holidays. Aldor-Noiman, Feigin, & Mandelbaum (2009) exclude holidays when training their model; however, we cannot ignore these days because some splits operate on holidays and may exhibit different behavior on those days. In order to capture this behavior, we include a holiday indicator (*holiday_ind*) by period-of-day interaction effect in the model. However, some training data sets may include only a single holiday, leading to high variance in the parameter estimates for this effect (each period observation from that one day becomes the new estimate for that period during holidays). To reduce the variability of these estimates, we combine groups of 3 periods together on holidays. That is, periods {1, 2, 3} are assigned *p_group* = 1, periods {4, 5, 6} are assigned *p_group* = 2, etc. The

*p_group*holiday_ind* interaction is included in the fixed effect structure as an additive effect.

Following Aldor-Noiman *et al*. (2009), we fit a linear mixed model with correlated errors to the transformed call counts
$$Y = X\beta + Zb + \varepsilon$$
where

- $Y$ is the vector transformed call counts, $Y = \sqrt{count + 0.25}$
- $X$ is a matrix containing the levels of the fixed effects for each observation
- $\beta$ is the vector of fixed effects parameters containing a day-of-week*period-of-day interaction and a *p_group*holiday-indicator interaction
- $Z$ is a binary coefficient matrix for the random day-to-day effects in the model. There is one column for each day in the data.
- $b \sim N(0, G)$ is the vector of random day-to-day effects. Each unique day in the data set is represented by one random effect in $b$. $G$ follows a first-order autoregressive structure, AR(1).
- $\varepsilon \sim N(0, R)$ is the vector of error terms (residuals), allowing $\varepsilon$ to potentially follow an AR(1) process within days. Thus, $R$ is a block-diagonal matrix, with one AR(1) block for each day in the data set. This accounts for the potential correlation in residuals from proximal periods within days.

The full model allows for complex correlation structures. However, for some splits (within particular training data sets), there may be only sporadic and sparse occurrences of call arrivals. This can lead to slow or failed model convergence in some cases. Aldor-Noiman *et al.* (2009) address this by estimating the doubly stochastic model in two steps: first, the inter-day correlation ($G$) is estimated using the aggregated total call counts from each day. These parameters are then held constant in a second call to SAS PROC MIXED while $\beta$ and $R$ are estimated.

Indeed, PROC MIXED can experience convergence problems when the solutions lie on the boundary of the parameter space, such as when variance components are zero (Karl, Yang, & Lohr, 2013). However, after making modifications to the default PROC MIXED settings, we were reliably able to achieve convergence of the full model with the joint optimization of $(\beta, G, R)$ in a single call to PROC MIXED. In this regard, our approach differs from that of Aldor-Noiman *et al.* (2009): we fit all of the model parameters jointly (with a single call to PROC MIXED). This will lead to reduced bias in the estimates for the models that do converge.

We improved convergence rates by changing the convergence criterion used by SAS PROC MIXED. By default, SAS ensures that the sum of squared parameter gradients (weighted by the current Hessian of the parameter estimates) is sufficiently small. However, in the presence of strong correlations in the doubly stochastic model, the parameter estimates may lie near the boundary of the parameter space, meaning the gradients may not approach 0 with convergence (Demidenko, 2013). As an alternative, we declare convergence when the relative change in the loglikelihood between iterations is sufficiently small. Additionally, we employ Fisher scoring during the estimation process. Fisher scoring is more stable for models with complex covariance structures and can lead to better estimates of the asymptotic covariance (Demidenko, 2013). Finally, since our application only uses the call volume point estimates and not the associated standard errors or tests of significance, we specify *ddfm=residual* to avoid spending substantial time calculating appropriate degrees of freedom for the approximate *F*-tests. If confidence or prediction intervals are needed, this value should be set to *ddfm=kenwardrodger2* in order to calculate Satterthwaite approximations for the degrees of freedom and to apply the Kenward-Rodger correction (Kenward & Roger, 2009) to the standard errors. The code for our modified approach appears in Figure 1.

```
proc mixed data=training_data scoring=50 maxiter=150 maxfunc=10000 convf=1E-6;
    class day_of_week period day_num split p_group;
    by split;
    /* The fixed effects */
    model transf_call_count=day_of_week*period p_group*holiday_ind/
        noint ddfm=residual outp=pred_call_count_output notest;
    /* The day-level random effects */
    /* Note: day_num_copy is not included in the CLASS statment and is numeric */
    random day_num / type=sp(pow)(day_num_copy);
    /* The period-level correlated residuals */
    repeated period / type=ar(1) subject=day_num;
run;
```

*Figure 1 Modified SAS code for the Doubly Stochastic Model*

The square root transformation is applied to reduce the right skew in the observed call volumes, and to stabilize the variance of the observations since quantities such as call volumes tend to follow a Poisson distribution. The approach in Figure 1 employs a normal approximation of this process. We experimented with fitting a mixed Poisson regression to the untransformed call volumes (via PROC GLIMMIX), but found that the run times became unfeasibly long (even when using the default pseudolikelihood approach and avoiding integral approximation) with no noticeable improvement in error rates.

**3. Choosing Recurrent Neural Network Configurations with a Designed Experiment**

Generally, neural networks consist of layers of weights and nonlinear activation functions that are used to relate inputs (predictors) to outputs (targets). Outputs from each layer are passed sequentially to the next layer as an input vector. The complexity of each layer is determined by the length of the output vector (number of nodes) it produces. A loss function is used to compare the output of the final layer of the neural network to the provided targets (e.g. call volumes), and an optimizer function provides updated values of the weights each node that will decrease the resulting loss. The "depth" of the model is controlled by the number of layers that are used. This "depth" is the source of the phrase "deep learning". For example, in image processing applications with convolutional neural networks, the different layers can be shown to represent different levels of granularity of detail in an image (Chollet & Allaire, 2018). Besides the number of layers and the number of nodes per layer, there are a number of choices that must be made regarding the properties of the optimizer, the distribution of the random initialization of the parameter weights, and the shape of the activation function(s).

In a traditional, densely connected network, the individual observations are assumed to be independent. A simple example using output from JMP Pro 14.1 helps to illustrate. Suppose we want to fit a densely connected neural network to predict the standardized call count using only the previous day's standardized call count at the same period (the lag-32 of the call count, since there are 32 periods per day in the example) as a predictor with one node in one layer, using a hyperbolic tangent activation function. This network shown in Figure 2 with resulting weights shown in Figure 3.

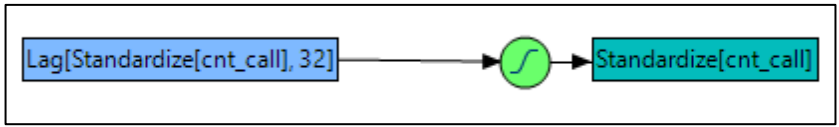

*Figure 2 Densely connected neural network with one node in one layer.*

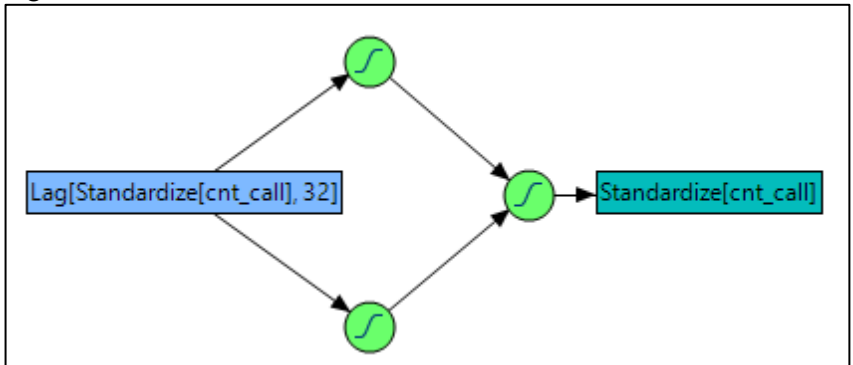

*Figure 3 Fitted weights from the network with one node in one layer.*

Suppose that the lag-32 standardized call count is equal to 1 at a given period, $t$. Then the neural network predicts a standardized call volume of

$$-0.6354 + 3.3923 * \text{TanH}(0.5 * (0.4046 + 0.5323 * 1)) = 0.847$$

for the current period, $t$, where *TanH* is the hyperbolic tangent function and the 0.5 parameter is a fixed value. The nonlinear activation function provides the network with the flexibility to model nonlinear relationships between the inputs and the response, as well as interactions between the inputs.

We next consider a slightly more complex network with two layers using 2 nodes in the first layer and 1 node in the second layer (Figure 4) with parameter estimates shown in Figure 5.

*Figure 4 A densely connected neural network with 2 layers using two nodes in the first layer and one node in the second layer.*

| Estimates | |
|---|---|
| Parameter | Estimate |
| H2_1:Lag[Standardize[cnt_call], 32] | -0.8785 |
| H2_1:Intercept | -0.3633 |
| H2_2:Lag[Standardize[cnt_call], 32] | 0.4124 |
| H2_2:Intercept | -0.0824 |
| H1_1:H2_1 | -0.7382 |
| H1_1:H2_2 | 0.3057 |
| H1_1:Intercept | 0.2416 |
| Standardize[cnt_call]_1:H1_1 | 4.9517 |
| Standardize[cnt_call]_2:Intercept | -0.8200 |

*Figure 5 Fitted weights from the network with two layers.*

Again assuming that the lag-32 standardized call count is 1, the predicted value is
$-0.82 + 4.9517$

$$* TanH\Big(0.5 \\ * \big(0.2416 - 0.7382 * TanH(0.5 * (-0.3633 - 0.8785 * 1)) + 0.3057 \\ * TanH(0.5 * (-0.0824 + 0.4124 * 1))\big)\Big) = 0.843$$

In the densely connected network, each observation is processed independently and there is no "memory" of what happened in the previously processed observation. In time series applications, however, there is a temporal ordering that the data are recorded in, and there may be correlation between nearby observations. For example, a spike or drop in call volume might persist over several periods. To address this potential, recurrent neural networks record information when fitting each observation that is then provided as a model input when fitting later observations.

In a simple (Elman) RNN layer, the output from each node (the output of the *TanH* functions, referred to as the "state") is recorded and stored, and used as an input for the same node when processing the next observation. Note that there is one state recorded for each node in the layer. For the single layer network example (Figure 3), the state was calculated as $s_t = \text{TanH}(0.5 * (0.4046 + 0.5323 * 1)) = 0.44$ when the lagged standardized call volume was equal to 1. A simple RNN learns an extra parameter (say, *u*) to act as a coefficient for the stored state, and the activation function $\text{TanH}(0.5 * (w_0 + w_1 * X_t))$ that is used by the dense network would be replaced by $s_t = \text{TanH}(0.5 * (w_0 + w_1 * X_t + u * s_{t-1}))$ in order to fit a simple RNN. The LSTM and GRU RNNs also use the recorded state when making predictions for the current time period, along with products of additional activation functions that are designed to carry state information further in time. Details of these additional structures are explained in Section 3 of Bianchi et al. (2018). For our purpose it is sufficient to note that the GRU network is extremely similar to the LSTM, albeit less complex due to the ommison of a group of paramters. Chollet & Allaire (2018) remark that Google Translate currently runs using an LSTM with seven large layers.

It is not clear *a priori* which of these four neural networks is most appropriate for a particular call center. Furthermore, it is possible that each of these networks may have a different optimal depth and structure when applied to the call center data. A designed experiment is run to identify the optimal model type and structure.

## 3.1 Data for the Experiment

We analyze call volumes aggregated into 30 minute periods from 3 different large-volume skills in an operational call center for the months of March-June 2018. All of the models under consideration are used to forecast next-day call volumes using 5 weeks of training data. Each day contains 32 30-minute periods during which the call center is operating. This application considers the Monday through Friday behavior of the call skills. Due to the use of the one-week lagged observations as a predictor in the neural networks, the first week of training data is not included in the predictor matrix (since the prior week's call volumes are unknown), meaning each training data set consists of 4*5*32=640 observations. The designed experiment in this section will evaluate the methods using a holdout period of the five one-day ahead predictions during last week in June using the three largest skills from the call center. Section 4 will then fit a reduced set of models over all skills for 60 day-ahead predictions.

## 3.2 Neural Network Input Factors

Each network makes use of the one-hot encoding (via a binary indicator matrix) of the day of week and the one-hot encoding of period of day. Furthermore, to capture the day- and week-long correlations, the networks are also fed the call volumes for the same period in the previous day when modeling the current day, as well as the call volumes for the same period on the same day of last week. One-period lagged call volumes are not included, as this is the purpose of the within-sequence memory of the RNNs. Other inputs include a binary indicator for whether the current day is a holiday, a binary indicator for whether yesterday was a holiday, a binary indicator for whether or not last week's observation was recorded on a holiday, and day number. The day number is a continuous counter for the number of the given day in the data set, which would potentially allow the neural network to detect trends across time. All told, these account for 42 vectors of input for the neural networks. There is no need to create indicator columns for interactions between day of week and period (or any other factors) as the neural network will automatically detect them.

Bianchi et al. (2017) application of hourly call volumes displays strong lag-24 correlation, representing a period-of-day effect. They remove this seasonality by differencing the data at lag-24. By contrast, we do not difference the call volumes, but instead include period-of-day (along with day-of-week) as exogenous variables and allow the neural network to detect this seasonality. This approach allows the network to detect the expected interactions between period-of-day and day-of-week, as well as any other input factors.

While these input vectors are included for all models, there are three final input vectors whose (joint) inclusion is treated as an experimental factor: the same-period predictions from the mixed model approach (Aldor-Noiman, Feigin, & Mandelbaum, Workload forecasting for a call center: Methodology and a case study, 2009), from a Winters smoothing model, and from a seasonal ARIMA$(1, 0, 1)(0, 1, 1)_{160}$ model. The inclusion of the predictions from these models as an input to the neural network is an original approach that gives the network the opportunity to form predictions that may be thought of as corrections to those from these traditional models, based on potential interactions with other included factors. For brevity, we refer to this as the mixed.cheat option, since it allows the neural networks to "cheat" by looking at the predictions generated by these other three models when forming its own predictions for the same time periods. If none of the other 42 input vectors were included with these three, this would represent a supervised learning approach to forming a dynamically weighted average of these three model predictions in order to create a single "bagged" prediction.

### 3.3 Network Configuration Aspects Treated as Experimental Factors

The designed experiment considered five different factors of structural settings of the neural networks: *model.type* {dense, simple (Elman) RNN, GRU, LSTM}, *nlayers* {1, 2}, *nnodes* (per layer) {25, 50, 75, 100}, *kernel.L2.reg* {0, 0.0001} and *mixed.cheat* {FALSE, TRUE}. The L2 regularizer adds *kernel.L2.reg* times each weight coefficient to the total loss function for the network. Similar to Lasso regression, this helps bound the magnitude of the model coefficients and could potentially help prevent overfitting the training data.

We fit a full factorial design (requiring 128 runs) for these five factors, which allows us to test for the presence of up to five-way interactions between the five factors. Figure 6 shows the first 6 runs of the design. The design is replicated over the 3 largest splits and across the 5 subsequent one-day ahead forecasts. This produces a total of 1920 runs for the entire experiment. The replication allows for behavior to be averaged over different days and for a more detailed exploration of how the variance of the error rates depends on each factor.

| model.type | nlayers | mixed.cheat | nnodes | kernel.L2.reg |
|---|---|---|---|---|
| layer_simple_rnn | 2 | FALSE | 50 | 0.0001 |
| layer_gru | 1 | FALSE | 25 | 0 |
| layer_lstm | 1 | TRUE | 100 | 0 |
| layer_gru | 1 | TRUE | 75 | 0.0001 |
| layer_gru | 1 | TRUE | 75 | 0 |
| layer_simple_rnn | 2 | FALSE | 100 | 0 |

*Figure 6 First 6 runs of the designed experiment*

### 3.4 Static Considerations for Neural Network Configuration

The input for the classical dense neural network is a 640x42 matrix (640x45 if *mixed.cheat=TRUE)*. The dense network does not consider the temporal ordering of the 640 observations (outside of the explicit inclusion of the day- and week-lagged observations as inputs): the observations are shuffled after each epoch and then processed in batches (we used a batch size of 32).

By contrast, the RNNs consider the ordering of the observations. The input for the RNNs is a 20x32x42 array. This indicates to the RNN that there are 20 batches (days) of 32 timesteps (periods) with 42 predictors per time step. By default, the batches are treated independently and the timesteps within each batch are potentially correlated (via the persistence of the states in the RNN). If the batches themselves are presented in a temporal order (as is the case in our application), then this can be indicated to the Keras model via the STATEFUL=TRUE option and by disabling the shuffling of batches during training. This retains the model weights from batch-to-batch (day-to-day) to allow for possible long-term behavior. However, we found that using the STATEFUL option led to a failure to converge in some *skill*day* combinations and the resulting validation error rates were not significantly different from those generated without the STATEFUL option. This seems to indicate that the dependence on prior days' behavior is already captured by the inclusion of the one-day and one-week lagged observations. Due to the occasional convergence issues, the results in the sequel are generated with STATEFUL=FALSE (and batch shuffling enabled during training). It would have also been possible to fit the model with week- or month-long batches by training the RNN on a 4x160x42 or a 1x640x42 array. We did not consider the week-long batches, but the month-long batches tended to produce inferior predictions to the day-long batches. This could possibly be due to the lack of long term correlations in the data and the fact that the smaller batches allow for the shuffling of the ordering that the days are fed through the gradient-optimization routine of the neural network, which can improve the model fit by preventing the model from overweighting the first observations that are provided to the network in each epoch (Chollet & Allaire, 2018).

Note that the model weights are reset and the model is retrained for each skill. This is in contrast to the approach taken by Zhu & Laptev (2017) in which a single network is fit to accommodate disparate behavior from different cities. As discussed in Section 5, a single multi-output network could potentially be built to model all of the skills at once.

While they were not included as experimental factors in this application, we also noticed a significant relationship between the quality of the predictions and the optimization routine employed. Extensive pilot experimentation led to our use of the AMSGrad variant (Reddi, Kale, & Kumar, 2018) of the Adam optimizer (Kingma & Ba, 2014), with a learning rate decay of 0.0001. We would recommend including both the optimizer and the optional learning rate decay as factors in the designed experiment for parameter tuning in future problems.

We found it was important to tune the number of epochs for each model fit using validation data (the last week in the training set) by first fitting 500 epochs for each model, taking a moving average (with a window size of 10 epochs) of the resulting WAPEs on the validation data, and then refitting the model (on both the training and the validation data in order to predict an additional day which was held out as a test set) with the number of epochs that produced the minimum validation WAPE. The moving average is important due to the volatile and non-monotonic behavior we observed in the individual recorded WAPEs and helps to find a relatively stable region. Consistent with previous findings (Bianchi, Maiorino, Kampffmeyer, Rizzi, & Jenssen, 2017), we noticed that the RNNs take many more epochs to converge than the dense neural network. Other researchers (such as Bianchi *et al.*) have used more than 500 epochs when fitting RNNs, so this upper bound should also be considered as an important factor when building an RNN.

While we also experimented with recurrent dropout (Gal & Ghahramani, 2015) to prevent overfitting, it led to degraded performance in the early iterations of our experiment and we removed it from consideration. However, this could simply be due to features of this particular application, such as the relatively short training period of five weeks: Chollet & Allaire (2018) strongly advocate the use of dropout and recurrent dropout.

The models experienced improved errror rates after switching the kernel initializer for the random weights to the He normal initializer (He, Zhang, Ren, & Sun, 2015). We used the relu activation function exclusively, although this choice could also impact the quality of the resulting model fit. And while our application did not detect any two-factor interactions among the five experimental factors, it is possible that some of these additional factors could depend on the type of (R)NN being used, meaning there would be an interaction between these factors and *model.type*.

A final contributing factor is the batch size. This determines how many observations are processed before the gradients are updated: when fitting Keras models on a GPU, amount of memory available on the GPU can be a limiting factor on the batch size. This choice is more constrained in the RNNs, where each batch is a single day/week/month (determined by the number of timesteps specified in the input array to the RNN). By contrast, the batch size for a dense network can be set between 1 and the number of observations in the training data. We noticed significant differences in the error rates from the dense model depending on what batch size was used.

### 3.5 Experimental Response
While mean squared error (MSE) is frequently used to evaluate and compare predictive models, this is a poor metric for the call center application as it will give undue focus to the low call volume periods at the beginning and end of each day. Instead, the weighted absolute percentage error (WAPE) is recommended for call volume modeling (Ibrahim, Ye, L'Ecuyer, & Shen, 2016). This weights the absolute percentage error in each period by the number of calls received in that period and is defined by

$$WAPE = \frac{\sum_{i=1}^{n} |Y_i - \hat{Y}_i|}{\sum_{i=1}^{n} Y_i}$$

where $Y_i$ and $\hat{Y}_i$ are the observed and predicted volumes, respectively, for each 30 minute period $i = 1, \ldots, n$. Because WAPE is our metric of interest, the models are compiled to use a mean absolute error loss-function, which minimizes the numerator of the WAPE (the denominator is static).

For each row of the experimental table (Figure 6), the specified neural network is fit (independently) to model five subsequent days of each split. That is, day 1 is predicted using the previous 5 weeks leading up to day 1, then the model is reset and day 2 is predicted using the 5 weeks leading up to day 2, etc. The vector of five one-day ahead predictions is compared with the observed call volumes (that were not visible to the model during training), and the resulting WAPE is recorded.

### 3.6 Analysis of Experimental Results
Figure 7 gives a typical output of the prediction error across the different formulations of the neural networks. GRU often has the lowest forecast error or at least is consistently close to the lowest. The other procedures tend to have much more unstable performance based on the choice of nodes and layers as well as across the days and splits.

|         |        | model.type |         |          |            |
|---------|--------|------------|---------|----------|------------|
|         |        | NN Classic | RNN_GRU | RNN_LSTM | RNN_Simple |
|         |        | WAPE       | WAPE    | WAPE     | WAPE       |
| nlayers | nnodes | Mean       | Mean    | Mean     | Mean       |
| 1       | 25     | 6.7%       | 6.2%    | 8.5%     | 6.0%       |
|         | 50     | 7.1%       | 6.2%    | 6.5%     | 6.7%       |
|         | 75     | 7.1%       | 5.8%    | 6.8%     | 5.9%       |
|         | 100    | 7.0%       | 6.4%    | 6.9%     | 6.8%       |
| 2       | 25     | 7.5%       | 6.6%    | 8.1%     | 6.9%       |
|         | 50     | 7.3%       | 5.8%    | 6.9%     | 7.0%       |
|         | 75     | 7.1%       | 6.0%    | 6.6%     | 6.5%       |
|         | 100    | 8.0%       | 5.9%    | 7.5%     | 6.0%       |

*Figure 7 Forecast error by model type and settings for forecast day 5 split 3*

Regression analysis is used to determine the statistically significant factors and interactions. In order to control a heavy right-skew in the recorded WAPEs, an inverse transformation is applied to use as the response in the regression models. Figure 12 displays the original skewness and the transformation to normality after taking the inverse of the WAPEs.

### 3.7 Loglinear Variance Regression Model

The mean structure of 1/WAPE is modeled by including the main effects of the five experimental factors (*model.type*, *nlayers*, *mixed.cheat*, *nnodes*, *kernel.L2.reg*) and all of the interactions. In addition, *split*, *file* (which represents the validation day), and *split\*file* are included as blocking factors.

While an ordinary regression model for the full factorial design would assume the same error variance for all responses across the design space, the loglinear variance model allows for the error variance itself to be modelled as a function of the input factors. This is appealing for this application, where parsimonious networks may be expected to demonstrate less variability during repeated fittings.

For each factor setting, the loglinear variance model produces a predicted mean of 1/WAPE and a predicted standard deviation of 1/WAPE, representing the variability due to the random initial weighting of the NN.

We include *model.type*, *nlayers*, *mixed.cheat*, *nnodes*, *kernel.L2.reg, file,* and *split* as factors in the loglinear variance model (using JMP Pro 14.1), which models the log of the error variance as a linear combination of these factors (which are all treated as categorical). With the exception of *kernel.L2.reg*, all of these effects are found to be significantly associated with the error variance (Figure 8).

**Variance Effect Likelihood Ratio Tests**

| Source | Test Type | DF | ChiSquare | Prob>ChiSq |
|---|---|---|---|---|
| model.type | Likelihood | 3 | 114.126 | <.0001* |
| nlayers | Likelihood | 1 | 4.6333 | 0.0314* |
| mixed.cheat | Likelihood | 1 | 53.6884 | <.0001* |
| nnodes | Likelihood | 3 | 16.5624 | 0.0009* |
| kernel.L2.reg | Likelihood | 1 | 0.0042 | 0.9482 |
| file | Likelihood | 4 | 36.8213 | <.0001* |
| split | Likelihood | 2 | 56.3123 | <.0001* |

*Figure 8 Factors with a significant impact on WAPE variability*

For the mean model, *model.type*, *nlayers*, *mixed.cheat*, *nnodes, day*, *split*, and *file\*split* were found to be significantly associated with 1/WAPE (Figure 9). Neither *kernel.L2.reg* nor any of the interaction terms involving the NN architecture were found to be significant.

| Fixed Effect Tests | | | | | |
|---|---|---|---|---|---|
| Source | Nparm | DF | DFDen | F Ratio | Prob > F |
| model.type | 3 | 3 | 796.5 | 44.6120 | <.0001* |
| nlayers | 1 | 1 | 1258 | 5.5213 | 0.0189* |
| model.type*nlayers | 3 | 3 | 796.5 | 1.0691 | 0.3614 |
| mixed.cheat | 1 | 1 | 1258 | 17.0652 | <.0001* |
| model.type*mixed.cheat | 3 | 3 | 796.5 | 1.3181 | 0.2672 |
| nlayers*mixed.cheat | 1 | 1 | 1258 | 2.3413 | 0.1262 |
| model.type*nlayers*mixed.cheat | 3 | 3 | 796.5 | 1.9528 | 0.1196 |
| nnodes | 3 | 3 | 786.2 | 4.5600 | 0.0036* |
| model.type*nnodes | 9 | 9 | 839.4 | 0.7864 | 0.6290 |
| nlayers*nnodes | 3 | 3 | 786.2 | 0.7371 | 0.5301 |
| model.type*nlayers*nnodes | 9 | 9 | 839.4 | 0.6668 | 0.7395 |
| mixed.cheat*nnodes | 3 | 3 | 786.2 | 0.6624 | 0.5753 |
| model.type*mixed.cheat*nnodes | 9 | 9 | 839.4 | 0.2224 | 0.9913 |
| nlayers*mixed.cheat*nnodes | 3 | 3 | 786.2 | 0.9291 | 0.4261 |
| model.type*nlayers*mixed.cheat*nnodes | 9 | 9 | 839.4 | 0.4312 | 0.9186 |
| kernel.L2.reg | 1 | 1 | 1258 | 0.0026 | 0.9590 |
| model.type*kernel.L2.reg | 3 | 3 | 796.5 | 0.1263 | 0.9445 |
| nlayers*kernel.L2.reg | 1 | 1 | 1258 | 0.2313 | 0.6306 |
| model.type*nlayers*kernel.L2.reg | 3 | 3 | 796.5 | 0.7812 | 0.5046 |
| mixed.cheat*kernel.L2.reg | 1 | 1 | 1258 | 0.0295 | 0.8636 |
| model.type*mixed.cheat*kernel.L2.reg | 3 | 3 | 796.5 | 0.1395 | 0.9364 |
| nlayers*mixed.cheat*kernel.L2.reg | 1 | 1 | 1258 | 0.1498 | 0.6988 |
| model.type*nlayers*mixed.cheat*kernel.L2.reg | 3 | 3 | 796.5 | 0.3776 | 0.7692 |
| nnodes*kernel.L2.reg | 3 | 3 | 786.2 | 0.9118 | 0.4347 |
| model.type*nnodes*kernel.L2.reg | 9 | 9 | 839.4 | 0.8905 | 0.5331 |
| nlayers*nnodes*kernel.L2.reg | 3 | 3 | 786.2 | 0.7442 | 0.5259 |
| model.type*nlayers*nnodes*kernel.L2.reg | 9 | 9 | 839.4 | 0.6535 | 0.7513 |
| mixed.cheat*nnodes*kernel.L2.reg | 3 | 3 | 786.2 | 0.0812 | 0.9703 |
| model.type*mixed.cheat*nnodes*kernel.L2.reg | 9 | 9 | 839.4 | 0.2691 | 0.9827 |
| nlayers*mixed.cheat*nnodes*kernel.L2.reg | 3 | 3 | 786.2 | 0.8510 | 0.4662 |
| model.type*nlayers*mixed.cheat*nnodes*kernel.L2.reg | 9 | 9 | 839.4 | 0.6614 | 0.7442 |
| file | 4 | 4 | 668.1 | 1258.316 | <.0001* |
| split | 2 | 2 | 857.3 | 1867.911 | <.0001* |
| file*split | 8 | 8 | 746.8 | 244.5502 | <.0001* |

*Figure 9 Factors that are significantly associated with the mean WAPE*

Removing the insignificant interaction terms (but allowing *kernel.L2.reg* to remain in both the mean and variance models) produces the profiler shown in Figure 10. Notice the large standard deviation associated with the LSTM model. The dependence on *file* and *split* is not shown, since there are no interactions modeled between these and the other experimental factors. While we want to **maximize 1/WAPE**, we also want to **minimize the standard deviation of 1/WAPE**. That is, we would like to maximize the lower 95% prediction interval of 1/WAPE (or minimize the upper 95% PI of WAPE) in order to find the model configuration that will give a combination of relatively good results on average while being protected against large errors.

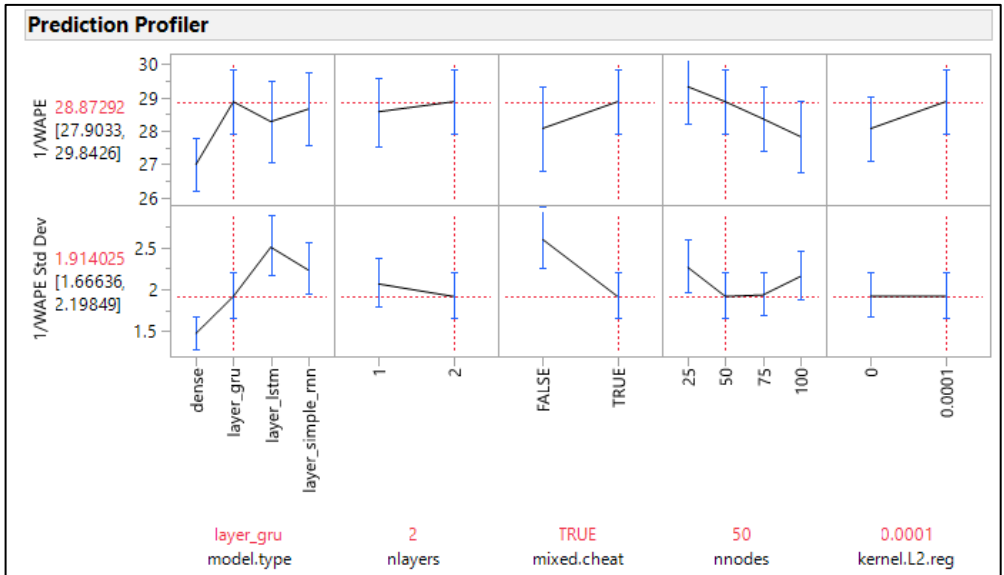

*Figure 10 Mean and Standard Deviation of 1/WAPE as a function of the experimental factors. Goal is to maximize the top (1/mean forecast error) and minimize the bottom (standard deviation)*

Figure 11 plots the upper 95% prediction interval for WAPE against the experimental factors. The model configuration that minimizes the upper 95% PI of WAPE is a GRU model that is allowed to use the MIXED, ARIMA, and Winters predictions as inputs, with 2 layers and 50 nodes per layer and an L2 penalty of 0.0001 on the kernel weights. However, the L2 penalty was not found to be significant in either the mean or the variance models and the contribution of *nlayers* is relatively flat.

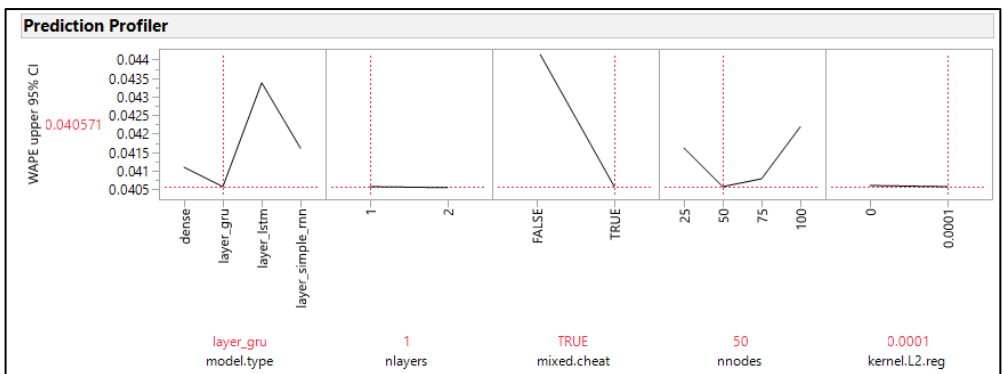

*Figure 11 Upper 95% Prediction Interval for WAPE*

To summarize, the designed experiment provided insight to effectively select the appropriate levels across several neural network models and parameters. The goal is to balance model performance in minimizing forecast error (that is, maximizing 1/forecast error) and minimizing the variance of this forecast error. Figure 10 and Figure 11 are interpretable across all factors and settings as displayed due to the absence of significant interactions between the factors. The top half of Figure 10 displays the reciprocal forecast error (the larger the value the better) which indicates preferred settings of GRU or Elman (simple) for the model, 1 layer, using the mixed model forecasts, and 25 nodes while the selection of L2.kernel makes little difference. The lower half of Figure 10 displays the variance (lower is better) where the traditional neural net would be preferred along with 2 layers, using mixed model forecasts, with 50 nodes while robust to the regularization parameter value. Note that the traditional dense neural network does have consistently worse forecast error across all scenarios and could not be recommended despite having the lowest variance. The prediction interval on forecast error is an alternative view that is preferred by many practitioners where the goal is to minimize the width. Figure 11 (lower is better) clearly shows GRU is the preferred solution using the mixed model forecasts with 50 nodes.

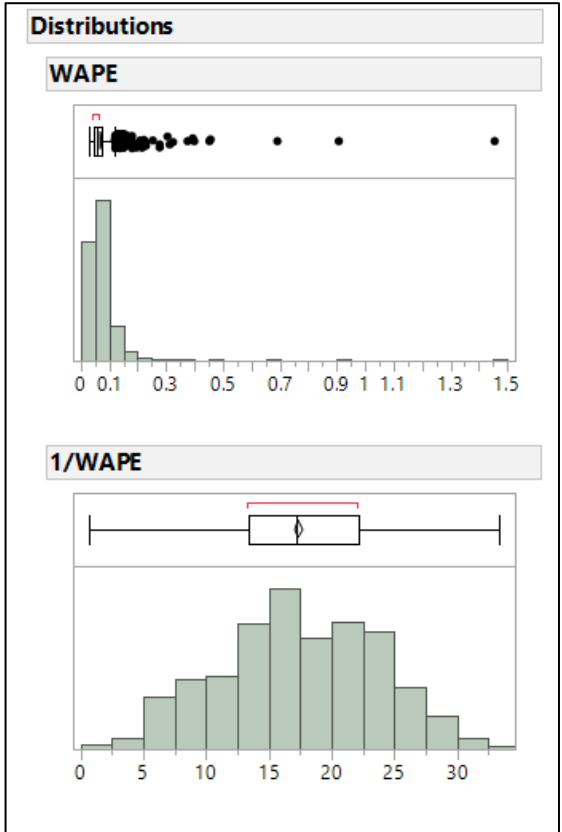

*Figure 12 Representative distribution of WAPE and 1/WAPE resulting from the designed experiment*

## 4. Comprehensive Performance Study

Based on the results of the designed experiment for NN error rates, we perform a further study with actual call center data across 36 skills rather than only 3. We consider a consistent 5-week training period advancing across multiple months of data to produce 60 one-day ahead predictions. These predictions do not use the actual data for the forecasted day during model training and can be viewed as a validation set for the trained models. This allows us to compare the relative performance of the mixed model and the neural networks. We also include the error rates from the ARIMA and Winters seasonal smoothing models for comparison.

Though we could have simply chosen the GRU RNN with 50 nodes on each of 2 layers with *kernel.L2.reg* complexity parameter set to 0.0001 as the only representative RNN based on the designed experiment, we decided to include all 4 neural network methods screened by using these same parameter settings. It is possible with the added skills – with many having much lower call volumes – that another RNN model could work better than GRU.

**4.1 Results: One-day ahead Predictions Over 60 Separate Validation Days**

Figure 13 provides the forecast errors averaged across all 60 days for the 12 splits with the highest call volume sorted in descending call volume order. Note that results in this section are generated by the (R)NNs with the *mixed.cheat* option disabled. The Doubly Stochastic Mixed Model has the lowest average forecast error for all but Split 5540 and is often significantly lower than the competitors. Winters Seasonal Exponential Smoothing also performs quite well given its relative simplicity. The GRU performance confirms the results from the designed experiment as usually the best RNN and always competitive with the best. The highly complex LSTM recurrent neural network has very large forecast errors for many of these splits and cannot be recommended. Note also that forecast error generally increases for all methods as call volume decreases.

| Split | Sum Call Vol | ARIMA | Doubly Stoch | NN_Classic | RNN_GRU | RNN_LSTM | RNN_Simple | Winters |
|---|---|---|---|---|---|---|---|---|
| 5000&5240 | 929754 | 9.0% | 8.4% | 10.2% | 11.2% | 38.0% | 13.9% | 8.3% |
| 5400&5570 | 461256 | 7.4% | 6.6% | 8.8% | 8.4% | 26.1% | 14.3% | 7.1% |
| 5620 | 162996 | 8.6% | 7.7% | 9.6% | 9.1% | 28.7% | 9.6% | 8.4% |
| 5660 | 137759 | 10.2% | 10.1% | 13.2% | 12.0% | 17.1% | 14.0% | 10.5% |
| 5020 | 71930 | 11.5% | 11.2% | 14.5% | 12.3% | 27.3% | 14.2% | 11.7% |
| 5630 | 65079 | 15.4% | 14.9% | 16.7% | 15.5% | 26.4% | 15.5% | 14.1% |
| 5840 | 63984 | 13.1% | 12.1% | 14.6% | 13.4% | 25.3% | 12.9% | 13.0% |
| 5200 | 48728 | 13.2% | 13.4% | 15.6% | 13.5% | 52.4% | 13.5% | 13.6% |
| 5540 | 38236 | 15.9% | 15.8% | 16.8% | 16.3% | 28.5% | 16.6% | 16.0% |
| 5670 | 34793 | 14.0% | 13.2% | 15.6% | 14.2% | 27.3% | 13.9% | 14.0% |
| 6500 | 30534 | 16.8% | 16.4% | 18.5% | 17.3% | 23.0% | 18.0% | 17.1% |
| 5260 | 23849 | 21.9% | 20.9% | 22.7% | 20.3% | 32.5% | 20.2% | 21.2% |

*Figure 13 Average WAPE forecast errors across 60 separate validation days for high call-volume splits*

Figure 14 shows the similar trend of increasing error rates with decreasing call volumes for the medium call volume splits. The GRU recurrent neural network is usually outperforming the other neural network methods and is closer to the error rates of the Doubly Stochastic Mixed Model.

| Split | Sum Call Vol | ARIMA | Doubly Stoch | NN_Classic | RNN_GRU | RNN_LSTM | RNN_Simple | Winters |
|---|---|---|---|---|---|---|---|---|
| 5460 | 20922 | 18.2% | 18.1% | 19.5% | 18.1% | 26.0% | 17.7% | 18.2% |
| 5410 | 19461 | 19.8% | 19.2% | 21.4% | 20.0% | 59.1% | 20.8% | 19.7% |
| 6350 | 16874 | 30.8% | 26.8% | 29.3% | 28.7% | 31.1% | 29.0% | 29.4% |
| 5060 | 16765 | 19.3% | 19.1% | 20.7% | 19.6% | 24.4% | 19.6% | 19.6% |
| 5650 | 9911 | 23.8% | 23.4% | 24.6% | 24.0% | 27.3% | 24.3% | 24.0% |
| 5030 | 8102 | 40.0% | 26.0% | 28.5% | 27.8% | 33.2% | 27.7% | 26.9% |
| 5680 | 7525 | 27.3% | 27.1% | 28.2% | 26.7% | 31.9% | 27.8% | 27.7% |
| 5440 | 7402 | 28.5% | 28.1% | 30.3% | 28.3% | 33.8% | 29.7% | 29.2% |
| 5070 | 6446 | 34.6% | 34.5% | 34.6% | 33.5% | 43.6% | 35.0% | 34.7% |
| 5420 | 5247 | 35.0% | 34.6% | 36.2% | 33.9% | 38.1% | 34.7% | 35.0% |
| 5899 | 4844 | 36.4% | 35.8% | 37.0% | 35.2% | 64.8% | 36.6% | 36.7% |
| 5100 | 4019 | 34.7% | 34.0% | 36.6% | 34.3% | 40.1% | 35.5% | 34.8% |

*Figure 14 Average WAPE forecast errors across 60 separate validation days for medium call-volume splits*

For the low call volume splits displayed in Figure 15, the GRU and Simple recurrent neural networks perform similarly and slightly better than Doubly Stochastic and Winters for most of the splits. The very low call volumes (last 3 splits) do seem to benefit from the recurrent neural network formulation.

| Split | Sum Call Vol | ARIMA | Doubly Stoch | NN_Classic | RNN_GRU | RNN_LSTM | RNN_Simple | Winters |
|---|---|---|---|---|---|---|---|---|
| 5710 | 3742 | 39.6% | 38.8% | 40.7% | 39.1% | 45.7% | 40.2% | 38.9% |
| 5820 | 2949 | 44.1% | 43.1% | 46.1% | 43.2% | 50.4% | 44.0% | 43.4% |
| 5690 | 2238 | 56.1% | 51.1% | 51.6% | 51.2% | 58.9% | 51.8% | 52.5% |
| 5220 | 2089 | 49.7% | 49.5% | 50.5% | 49.3% | 54.0% | 50.1% | 49.7% |
| 5470 | 1975 | 49.2% | 48.8% | 52.0% | 50.5% | 57.0% | 50.7% | 50.2% |
| 6330 | 1556 | 60.5% | 60.1% | 61.0% | 60.2% | 65.1% | 61.6% | 60.7% |
| 5040 | 1398 | 79.0% | 69.0% | 69.8% | 73.4% | 72.2% | 71.1% | 80.7% |
| 6310 | 922 | 98.5% | 92.0% | 92.2% | 88.1% | 91.0% | 89.5% | 92.8% |
| 5720 | 918 | 80.2% | 77.8% | 77.5% | 77.6% | 85.0% | 78.4% | 80.1% |
| 6370 | 485 | 95.3% | 93.6% | 96.5% | 91.3% | 111.3% | 94.0% | 95.2% |
| 6360 | 171 | 150.4% | 136.8% | 126.5% | 107.7% | 118.7% | 111.7% | 142.5% |
| 6340 | 14 | 189.7% | 161.4% | 188.2% | 102.7% | 108.9% | 107.8% | 187.7% |

*Figure 15 Average WAPE forecast errors across 60 separate validation days for low call-volume splits*

Overall, the best performing procedures were the Doubly Stochastic Mixed Model and the GRU Recurrent Neural Network. Figure 16 shows forecast error by split sorted by call volume. Generally, the mixed model does better for large and medium volume splits (lower is better on the graph) while the RNN
model is more effective for the small volume—particularly the very small volume splits.

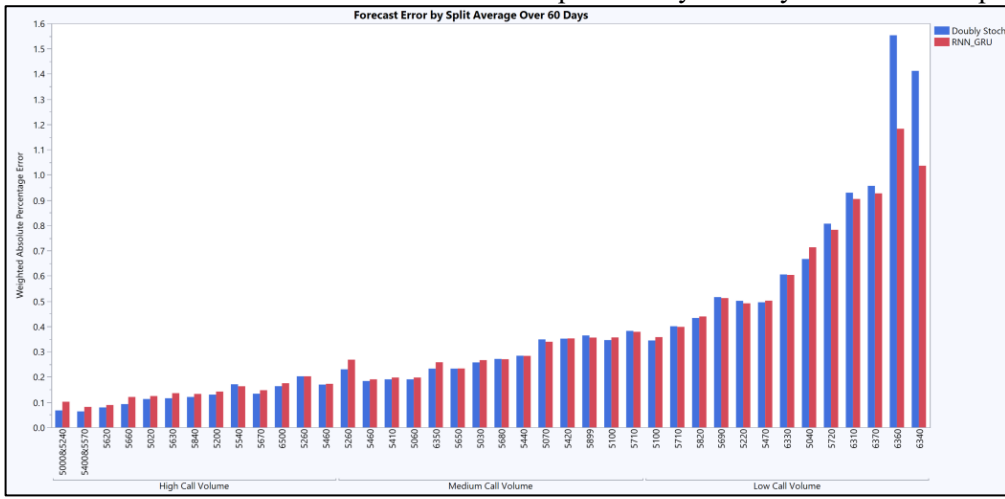

*Figure 16 Forecast error rates ordered by call volume by split for Doubly Stochastic (blue) and GRU (red)*

Figure 17 presents a different view of this same pattern. For each of the 60 one-day ahead forecasts within each split, the GRU and Doubly Stochastic WAPEs are recorded, along with the number of calls recorded for that split over the training and validation data (*sum_all_calls*). For each split, Figure 17 plots the percent of the 60 day-ahead forecasts for which GRU "won" (GRU WAPE < Doubly Stochastic WAPE) against the log of the median call volume recorded by that split over the 60 different pairs of training and validation data. There appears to be a linear decrease of the relative performance of GRU against Doubly Stochastic in the log of the call volume.

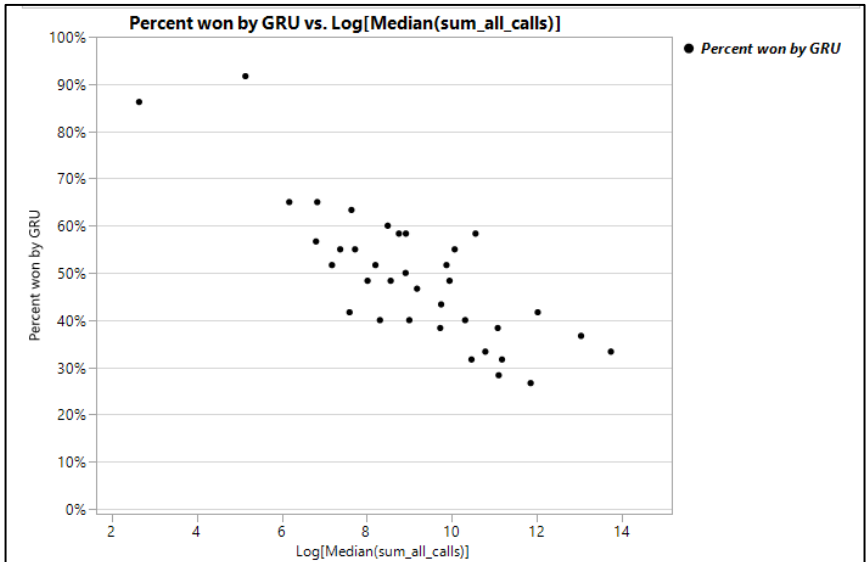

*Figure 17 Percent of the 60 day-ahead forecasts for which the GRU WAPE was less than the Doubly Stochastic WAPE for each of the 36 splits plotted against the log of the median sum of all calls recorded for each split over each pair of training and validation data.*

For the large call-volume splits, the extra flexibility of the GRU model does not lead to improvements over the predictions generated by the mixed model approach. Because the mixed model computations are faster and the implementation is less complex, there does not appear to be any benefit to running the neural networks for the high-volume splits for this short-term application. This is consistent with the findings of the Uber traffic volume study when short-term predictions were considered (Zhu & Laptev, 2017), and would possibly change if a longer training period (several months or multiple years) were used for the call center data.

**4.2 Improving GRU RNN by Using Doubly Stochastic Forecasts as a Covariate**
Based on pilot studies during the designed experiment, the forecasting performance of the GRU recurrent neural network often improved by integrating the forecasted value for the validation days from the doubly stochastic, ARIMA, and Winters models. This "cheating" by using other models' forecasts (shown as *mixed.cheat*) proved to be a significant benefit for the GRU model over these 60 predictions for each skill. The WAPE for the held-out validation data was again the primary measure of performance.

Figure 18 displays the forecast errors for the high call volume skills averaged over the 60 validation days for the each of the neural networks when they use the mixed forecasts. Note the side-by-side comparison of the RNN_GRU (no cheat) and GRU_cheat columns where in most cases the "cheating" does result in improved forecasts and in those cases where it is not better, it has only marginally declined. Additionally, the Simple_cheat error rates compare favorably with the GRU_cheat while the LSTM_cheat suffers from significantly poorer performance and instability issues. The doubly stochastic forecast is still quite good and often the best choice. These results are also consistent when looking at the medium and low call volume splits. Therefore, we recommend using the forecasts from a doubly stochastic or Winters model as inputs to recurrent neural networks.

| Split | Sum Call Vol | Doubly Stoch | RNN_GRU | GRU_cheat | LSTM_cheat | Simple_cheat |
|---|---|---|---|---|---|---|
| 5000&5240 | 926493 | 6.7% | 10.2% | 7.8% | 14.4% | 7.8% |
| 5400&5570 | 459961 | 6.3% | 8.1% | 7.4% | 9.4% | 6.9% |
| 5620 | 164138 | 7.9% | 8.9% | 8.8% | 15.2% | 8.7% |
| 5660 | 138389 | 9.2% | 12.0% | 9.9% | 18.1% | 9.8% |
| 5020 | 71900 | 11.2% | 12.4% | 12.4% | 27.0% | 12.3% |
| 5630 | 65279 | 11.5% | 13.5% | 13.7% | 421.5% | 13.2% |
| 5840 | 63687 | 12.0% | 13.2% | 13.1% | 26.1% | 12.7% |
| 5200 | 48624 | 13.0% | 14.2% | 14.4% | 28.1% | 14.0% |
| 5540 | 38318 | 17.1% | 16.3% | 16.9% | 20.8% | 17.5% |
| 5670 | 34741 | 13.4% | 14.7% | 14.4% | 24.5% | 14.2% |
| 6500 | 30352 | 16.3% | 17.5% | 16.8% | 35.2% | 17.2% |
| 5260 | 23674 | 20.7% | 21.3% | 20.5% | 28.0% | 20.8% |

*Figure 18 Forecast errors for high call volume splits averaged over 60 validation forecast days allowing NN to "cheat" to improve predictions*

The median of the GRU WAPE (*mixed.cheat*=FALSE) minus the GRU WAPE (mixed.cheat=TRUE) (within each split/day combination, for a sample size of 36*60=2160) is 0.002 with a p-value of 0.0005 from the Wilcoxon signed rank test with a null hypothesis that the differences were drawn from a population with median equal to 0, indicating that *mixed.cheat* does tend to improve the performance of the GRU model.

A similar comparison of paired differences of error rates (all with mixed.cheat=FALSE) confirms that GRU outperforms the other (R)NNs: LSTM - GRU produces a median of 0.0247 with a p-value of 1e-129, NN Classic - GRU produces a median of 0.0690 with a p-value of <1e-185, and Simple RNN - GRU produces a median of 0.0021 with a p-value of 1e-04.